\documentclass[aps,prl,twocolumn,superscriptaddress,floatfix,nofootinbib]{revtex4-1}

\usepackage{aas_macros}
\usepackage{graphicx,color,amsmath,amssymb,amsfonts,mathtools}
\usepackage[dvipsnames]{xcolor}
\usepackage[normalem]{ulem} 

\usepackage[colorlinks]{hyperref}
\hypersetup{
    colorlinks=true,
    linkcolor=NavyBlue,
    filecolor=NavyBlue,      
    urlcolor=NavyBlue,
    citecolor=NavyBlue,
    pdfpagemode=FullScreen
}

\newcommand{\orcid}[1]{\hspace{1mm}\href{https://orcid.org/#1}{\includegraphics[height=0.3cm,keepaspectratio]{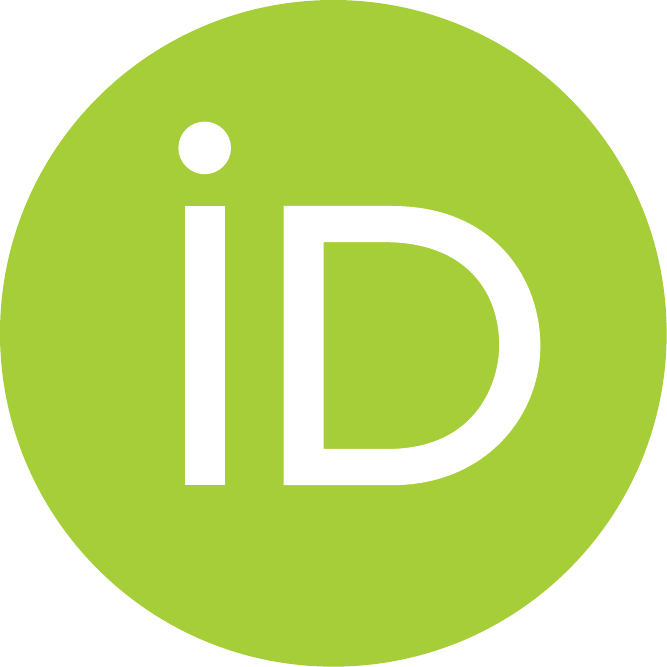}}}

\begin{document}
\title[Signature of Energy Losses on the Cosmic Ray Electron Spectrum]{Signature of Energy Losses on the Cosmic Ray Electron Spectrum}

\author{Carmelo Evoli\orcid{0000-0002-6023-5253}} 
\email{carmelo.evoli@gssi.it}
\affiliation{Gran Sasso Science Institute (GSSI), Viale Francesco Crispi 7,
67100 L'Aquila, Italy}
\affiliation{INFN-Laboratori Nazionali del Gran Sasso (LNGS), via G.~Acitelli
22, 67100 Assergi (AQ), Italy}

\author{Pasquale Blasi\orcid{0000-0003-2480-599X}}
\email{pasquale.blasi@gssi.it}
\affiliation{Gran Sasso Science Institute (GSSI), Viale Francesco Crispi 7,
67100 L'Aquila, Italy}
\affiliation{INFN-Laboratori Nazionali del Gran Sasso (LNGS), via G.~Acitelli
22, 67100 Assergi (AQ), Italy}

\author{Elena Amato\orcid{0000-0002-9881-8112}}
\email{amato@arcetri.astro.it}
\affiliation{INAF, Osservatorio Astrofisico di Arcetri, Largo E.~Fermi 5 -
I-50125 Firenze, Italy}

\author{Roberto Aloisio\orcid{0000-0003-0161-5923}}
\email{roberto.aloisio@gssi.it}
\affiliation{Gran Sasso Science Institute (GSSI), Viale Francesco Crispi 7,
67100 L'Aquila, Italy}
\affiliation{INFN-Laboratori Nazionali del Gran Sasso (LNGS), via G.~Acitelli
22, 67100 Assergi (AQ), Italy}

\begin{abstract}
We show that the fine structure of the electron spectrum in cosmic rays, especially the excess claimed by AMS-02 at energies $\gtrsim 42$ GeV, is fully accounted for in terms of inverse Compton losses in the photon background dominated by ultraviolet, infrared, and CMB photons, plus the standard synchrotron losses in the Galactic magnetic field. The transition to the Klein-Nishina regime on the ultraviolet background causes the feature. Hence, contrary to previous statements, observations do not require the overlap of different components. We stress that the feature observed by AMS-02 at energies $\gtrsim 42$ GeV is not related to the positron excess, which instead requires the existence of positron sources, such as pulsars. Because energy losses are the physical explanation of this feature, we indirectly confirm that the transport of leptons in the Galaxy is loss dominated down to energies of the order of tens of GeV. This finding imposes strong constraints on the feasibility of alternative theories of cosmic transport in which the grammage is accumulated in cocoons concentrated around sources, requiring that electrons and positrons become loss dominated only at very high energies. 
\end{abstract}

\maketitle

{\it Introduction.} The precision measurement of cosmic ray (CR) spectra carried out by AMS-02 onboard the International Space Station is profoundly affecting our views on the origin of CRs. The recent measurement of the electron spectrum \cite{AMS02-electrons} up to $\sim$TeV energies has revealed a surprising feature arising at $42.1^{+5.4}_{-5.2}$ GeV, consisting in a smooth hardening. This feature is not related to the rising positron fraction, as shown by the precision measurement of the positron spectrum by AMS-02 \cite{AMS02-positrons}. Data analysis of this feature has led to the conclusion that it can be fitted by assuming that two electron components overlap, the first with a very steep spectrum, corresponding to a slope $\gamma=-4.31\pm0.13$ and another with slope $\gamma=-3.14\pm 0.02$ \cite{AMS02-electrons}. The first of these components seems to have characteristics that are at odds with any known type of sources of astrophysical origin. 

Here we show that no such additional component is required and that in fact the feature at $\sim 40$ GeV arises naturally when inverse Compton scattering (ICS) off the photons populating the interstellar medium (ISM) is properly treated. Such background is made of several components ranging from the microwaves (CMB) to the IR, to optical and up to the ultraviolet (UV). The latter is actually the dominant photon background (in energetic terms) at high frequency. Electrons propagating in the ISM lose energy by scattering off the background light through ICS. The cross section for ICS is basically the Thompson cross section $\sigma_{T}$ as long as the scattering occurs on photons with energy $\epsilon$ such that $E\epsilon\ll m_{e}^{2} c^4$, where $m_e$ is the electron mass and $E$ is the electron energy in the lab frame. When this condition is not fulfilled, the scattering occurs in the Klein-Nishina (KN) regime and the corresponding cross section is correspondingly reduced. While this transition has little impact on electron losses when the scattering is dominated by CMB, infrared (IR) and optical light, the situation changes when UV photons are included. 
The typical temperature range corresponding to such photons is $8 \times 10^3  \lesssim T \lesssim 3\times 10^{4}$~K~\cite{Popescu2017}; hence, the KN effects become important at $E \simeq \frac{m_{\rm e}^{2} c^4}{2k_{\rm B}T_{\rm UV}}\sim 50$ GeV, although the effect in the rate of energy losses is already visible at somewhat lower energies. 
When the electron energy is much larger, losses become dominated by Thompson ICS scattering on the CMB and synchrotron emission in the Galactic magnetic field. If the electron transport is loss dominated, which is the case for electron energies greater than or similar to few GeV, as we show below, this transition reflects on the spectrum of leptons as a feature that has the same characteristics as the one observed by AMS-02. In principle, such a feature would also be present in the positron spectrum. However the spectrum of positrons in the energy region $\gtrsim 10$ GeV is an overlap of secondary positrons produced in inelastic pp collisions and the contribution that is typically associated to pulsars~\cite{Amato2018,Manconi2019,Recchia2019,Fornieri2020}, so that the feature is harder to spot in the positron spectrum. We prove that the AMS-02 feature is not due to the presence of electrons from pulsars and cannot reflect the energy dependence of the diffusion coefficient that needs to be invoked to explain the hardening observed in the spectra of nuclei~\cite{Genolini2017,Evoli2019}. In general, features in the electron spectrum would reflect in the diffuse radio emission~\cite{Orlando2013,Orlando2018} although the weakness of this feature and the broadness of the synchrotron kernel make it unobservable with current radio data.

We emphasize that the detection of this feature is the best evidence so far that electron transport in the Galaxy is dominated by energy losses, thereby casting doubt on alternative models of CR transport requiring a very short escape time of leptons \cite{Cowsik2016,Lipari2017,Lipari2019}. In fact, in such models, energy losses become important only at energies above a few hundred GeV. 

{\it CR lepton propagation in the Galaxy.} The simplest description of cosmic ray propagation in the Galaxy, in the presence of energy losses, is provided by the diffusion-loss equation ~\cite{Berezinskii1990}:
\begin{multline}\label{eq:transport}
\frac{\partial}{\partial t} n_e(t, E, \vec r) = D(E) \nabla^2 n_e(t, E, \vec r)
\\ - \frac{\partial}{\partial E}\left[ b(E) n_e(t, E, \vec r) \right] + {\cal
Q}(t, E, \vec r),
\end{multline}
where $n_e(\vec r,t,E) = dN/dVdE$ is the isotropic part of the differential CR lepton density, related to the differential flux as $\Phi = (d^4N)/(dEdAdtd\Omega) = n_e c/ 4\pi$. In Eq. \ref{eq:transport} we assumed that transport is mainly diffusive, with a diffusion coefficient $D(E)$ that is taken to be spatially constant. Since below we focus on energies $\gtrsim 10$ GeV, we ignore the effect of advection and possible second order reacceleration. Energy losses are described by the rate $b(E) \equiv dE/dt$, for particles of given energy $E$. The injection rate, discussed below, is described through the function ${\cal Q}(\vec r,t,E)$ in Eq.~\ref{eq:transport}. As usual, Eq.~\ref{eq:transport} is solved with the standard free-escape boundary condition at $|z|=H$ namely $n_{e}(|z|=H)=0$.
The diffusion coefficient in the Galaxy can be derived from observations of the ratios of fluxes of secondary and primary nuclei. This information leads to fit the quantity $H/D(E)$, where $H$ is the size of the halo and here we adopt the same $D(E)/H$ as derived in Ref.~\cite{Evoli2019}. A similar investigation was carried out including unstable isotopes, such as $^{10}$Be in Ref.~\cite{Evoli2020}, where the conclusion was reached that relatively large halos are preferred, $H\gtrsim 5$~kpc (here we assume $H=5$~kpc).  

Given the potentially important role of energy losses for high energy leptons, the stochastic nature of the sources needs to be taken into account, as discussed in Refs.~\cite{Atoyan1995,Mertsch2011,Blasi2012b,Pohl2013}. This purpose is most easily accomplished by adopting a Green function formalism. 
The contribution to the lepton spectrum due to an individual source $i$ active at time $t_s$ and at Galactic location $\vec r_s$ is provided by the Green function of the transport equation, hence the flux of cosmic leptons that are observed at the Sun position ($t_\odot$, $\vec r_\odot$) at an energy $E$ from that source can be written as:
\begin{equation}\label{eq:burstsolution}
\Phi_i(t_\odot, E, \vec r_\odot) = \frac{c}{4\pi}\frac{Q(E_*)b(E_*)}{(\pi \lambda^2_*)^{1/2}} \mathcal{G}_{\vec r}(|\vec r - \vec r_\odot|, E, E_*)
\end{equation}
where $\mathcal{G}_{\vec r}$ is the spatial part of the Green function that satisfies the free-escape boundary condition at $z = \pm H$~\cite{Delahaye2010} and $Q(E_s)$ is the source injection spectrum $dN/dE$. 
Here a particle injected with energy $E_s$ is observed after a time $\Delta t \equiv t_\odot - t_s$ with energy $E < E_s$ only if the elapsed time corresponds to the average time during which the energy of a particle decreases from $E_s$ to $E$ due to losses. Therefore $E_*$ is obtained by inverting the equation $t_s - t_\odot - \Delta \tau(E_*, E_s) = 0$, where the loss time is defined as:
$\Delta \tau(E, E_s) \equiv \int_E^{E_s} \frac{dE'}{|b(E')|}$.
In Eq.~\ref{eq:burstsolution} we introduced the propagation scale $\lambda_e$ which characterizes the lepton horizon, namely the maximum distance from which an electron of given energy can reach Earth propagating diffusively under the action of energy losses:
\begin{equation}\label{eq:lambda}
\lambda_e^2(E, E_s) \equiv 4 {\int_E^{E_s} dE' \, \frac{D(E')}{|b(E')|}}.
\end{equation}

For electrons and positrons with energy above a few GeV the main channels of energy losses while propagating in the Galaxy are inverse Compton scattering off the interstellar radiation field (ISRF) and the cosmic microwave background, and synchrotron emission in the Galactic magnetic field. The rate of energy losses can then be written as:
\begin{equation}\label{eq:losses}
b(E) = -\frac{4}{3} c \sigma_T \left[ f_{\rm KN}(E) U_\gamma + U_B \right] \left(\frac{E}{m_e c^2}\right)^2 \, , 
\end{equation}
where $\sigma_{T}$ is the Thompson scattering cross section and $U_i$ denotes the field energy densities. The function $f_{\rm KN}$ (see below) describes deviations of the ICS cross section from $\sigma_{T}$ (Klein-Nishina regime). Other mechanisms, such as bremsstrahlung and ionization losses become important at lower energies, that are not discussed here. 
We assume U$_{\rm B} = 0.25$~eV~cm$^{-3}$ (corresponding to a magnetic field B$_{0} \sim 3$~$\mu$G), and a multicomponent photon field made of the interstellar radiation field and CMB. 
The ISRF, as provided by Ref.~\cite{Delahaye2010}, has been obtained by fitting several blackbody spectra against the ISRF model distributed with the broadly used GALPROP code~\cite{Moskalenko2006} after averaging it over a cylinder of half height and radius of 2 kpc. As a result, it has been found that the local ISRF can be well approximated with 5 blackbody spectra corresponding to the infrared ($\rho_{\textrm{IR}}= 0.25$~eV/cm$^3$, $T_{\textrm{IR}} = 33.07$~K), optical ($\rho_\star = 0.055$~eV/cm$^3$, $T_\star = 313.32$~K), and 3 UV ($\rho_{\textrm{UV}} = 0.37,0.23,0.12$~eV/cm$^3$, $T_{\textrm{UV}} = 3249.3, 6150.4, 23209.0$~K) backgrounds. We additionally checked that our results remain unchanged by adopting a more recent derivation of the ISRF as, for example, the one presented in Ref.~\cite{Popescu2017}.

\begin{figure}[t]
\centering
\hspace{\stretch{1}}
\includegraphics[width=0.48\textwidth]{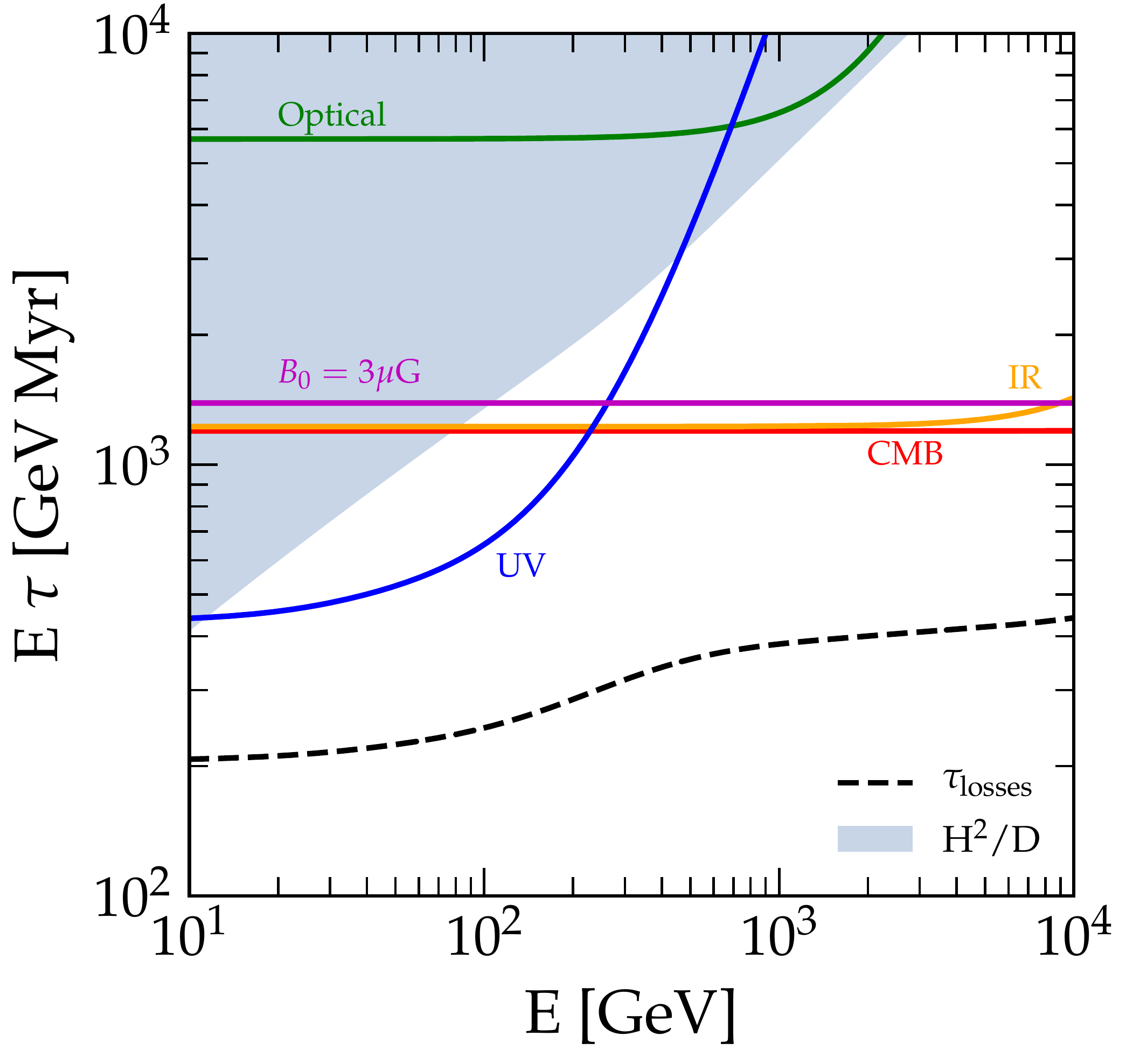}
\hspace{\stretch{1}}
\caption{Energy loss timescale as a function of the energy of CR electrons during their propagation in the Galaxy. The timescales are multiplied by $E$ to give prominence to the deviations from the standard $b \propto E^2$ regime. The dashed line represents the total timescale, while the solid lines refer to the single contributions by the magnetic field (magenta line) or ISRF components. The shadow region marks out the escape timescale from the Galaxy due to diffusion.}
\label{fig:losses}
\end{figure}

For a blackbody spectrum corresponding to a temperature $T_{i}$, the cross section for ICS enters the KN regime at energy $E = m^2_e c^4/2 k_B T_i$, and the cross section is modified with respect to the Thompson value as described by the approximated correction factor \cite{Hooper2017}:
\begin{equation}\label{eq:fkn}
f_{\rm KN}(E) \approx \frac{45 / 64 \pi^2 (m_e c^2/ k_{\rm B}T_i)^2}{45 / 64 \pi^2 (m_e c^2/ k_{\rm B} T_i)^2 + (E^2 / m_e^2 c^4)},
\end{equation}
where $k_{B}$ is the Boltzmann constant. The competition between energy losses and diffusion can be illustrated by comparing the timescales for the two processes, as shown in Fig.~\ref{fig:losses}. The energy loss timescale is defined as $\tau_l \sim E / b(E)$, as a function of lepton energy, while the timescale of diffusive particle escape is $t_d = H^2 / D(E)$.

Figure~\ref{fig:losses} shows several important facts: 1) Energy losses dominate electron transport at all the energies of interest here. Clearly this implication would become stronger for larger values of the halo size $H$, still allowed by the observed Be/B ratio~\cite{Evoli2020}. 2) At energy $\sim 50$~GeV, the ICS losses enter the KN regime with respect to the UV photon background ($T_i \lesssim 3 \times 10^4$~K); namely the cross section gets substantially reduced compared with $\sigma_{T}$. This phenomenon regulates the transition to CMB as the dominant photon background (plus the synchrotron contribution, present at all energies). In that regard energy losses are never in the full KN regime where they cannot be described as continuous (as in Eq.~\ref{eq:transport}). 3) The transition to the KN regime for the IR, the optical and CMB backgrounds has negligible effects on the transport of leptons. 

As a result of these well-established pieces of physical information, the total timescale for losses has a pronounced feature that starts at $\sim 40$ GeV and is due to the KN transition on the UV background. 

\begin{figure*}[t]
\hspace{\stretch{1}}
\includegraphics[width=0.49\textwidth]{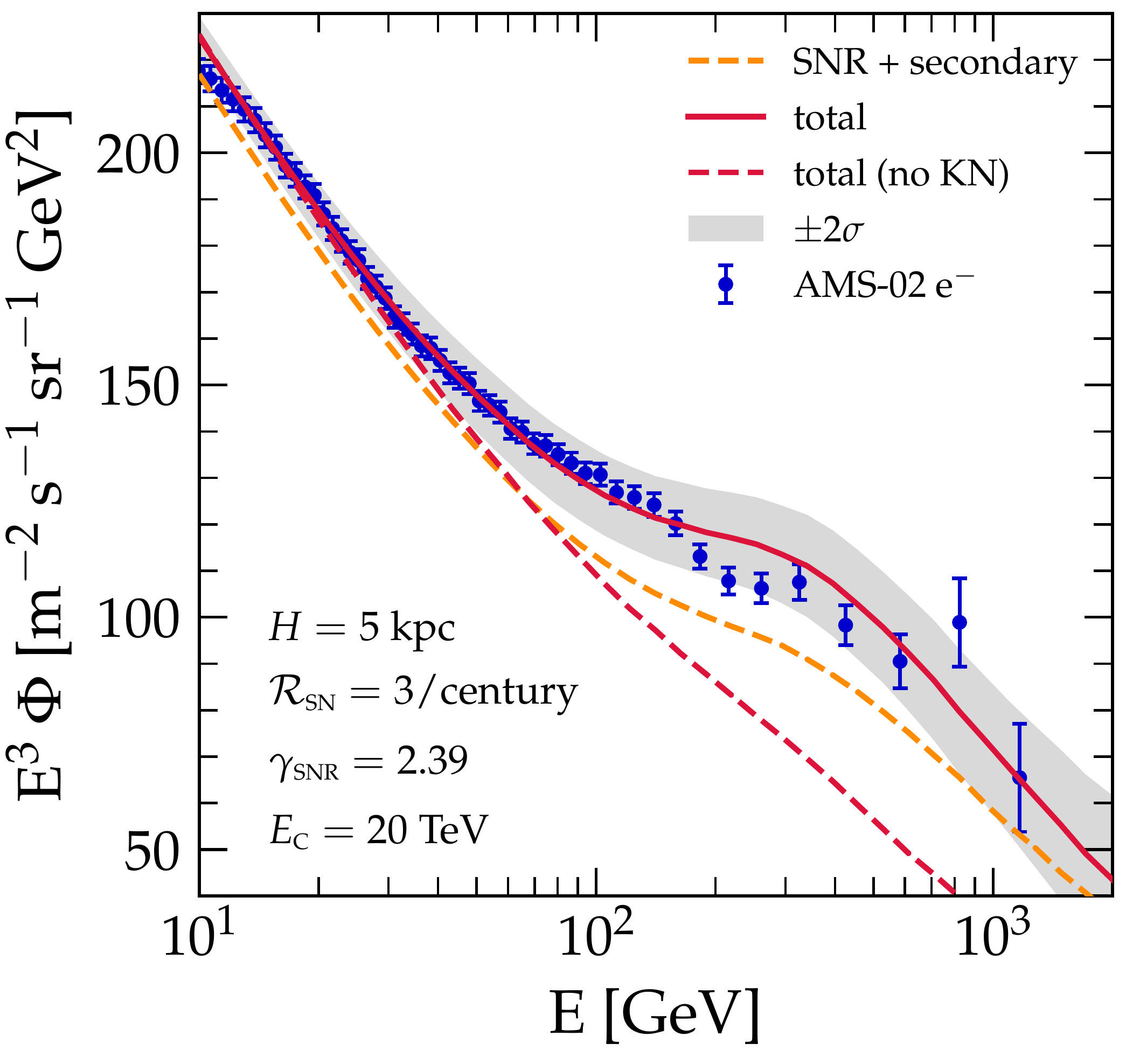}
\hspace{\stretch{2}}
\includegraphics[width=0.49\textwidth]{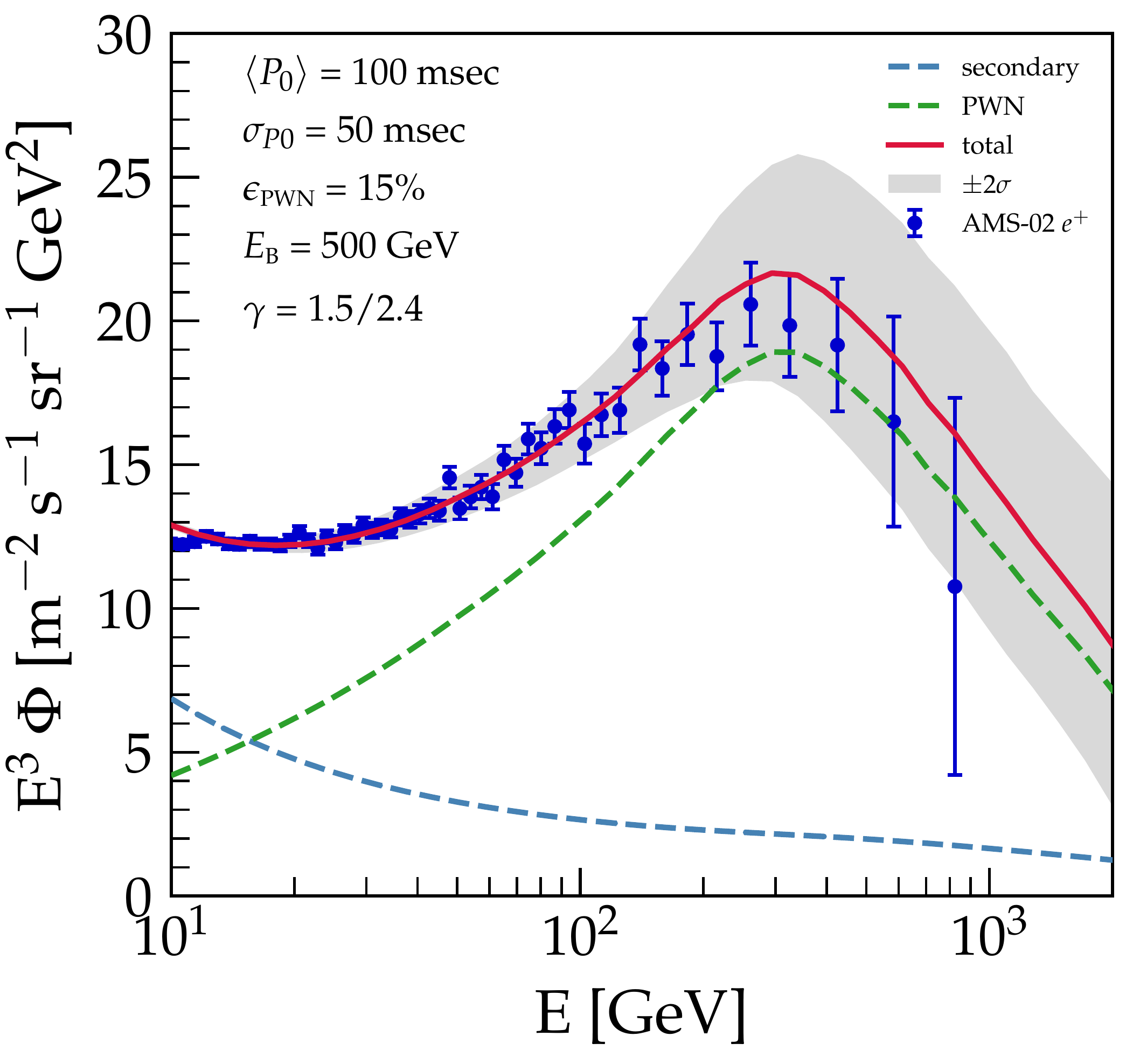}
\hspace{\stretch{1}}
\caption{The spectrum of cosmic ray electrons (left) and positrons (right) resulting from the sum of all sources. We also show the prediction for secondary leptons (blue dashed line) and positrons from PWNe (green dashed line). In the left panel, the sum of primary electrons from SNRs and secondaries (dashed orange line) and the total flux obtained neglecting the KN effect (dashed red line) are also shown.}
\label{fig:spectra}
\end{figure*}

{\it Results.} Electrons in the cosmic radiation are produced by sources of primary CRs, as a result of CR interactions in the ISM and finally by the potential sources of positrons, required by the observation of a rising positron fraction. We assume that primary electrons are accelerated at SNR shocks, located in time and space in a stochastic way in the Galaxy, as in Refs.~\cite{Blasi2012b,Amato2018}. The distribution of SNRs is determined by modeling the Galaxy with the four-arm model with logarithmic spiral arms taken from~\cite{SteimanCameron2010}, and weighing it in such a way that the distribution in galactocentric radius agrees with the distribution provided by Ref.~\cite{Green2015}. SNe are generated at a rate $\mathcal R$ = 3/century, assumed to remain constant over the longest timescales of CR confinement in the Galaxy, $\mathcal{O}$(100 Myr). The injection spectrum of CR electrons at a SNR shock is assumed to be described by a power law with a superexponential cutoff (specific for the case of Bohm diffusion in the acceleration region)~\cite{Zirakashvili2007,Blasi2010},
%
%
where the normalization $Q_0$ and the injection slope of this primary component $\gamma$ are fitted to the local observed spectrum. The cutoff energy $E_c$ is set by equating acceleration and losses timescales in the accelerator, and for typical conditions one gets $E_c \sim 10-100$~TeV~\cite{Vink2012}, and we assume for the sake of definitiveness that $E_c=20$~TeV. 

Following Ref.~\cite{Amato2018}, we also consider a second population of electrons and positrons injected (in equal amounts) by bow-shock PWNe, formed when pulsars associated to core collapse SN explosions (about 80\% of the total) leave the parent remnant and move into the ISM. The escape time of the pulsar from the remnant is calculated by assigning to each pulsar a birth kick velocity according to the distribution provided by Ref.~\cite{FaucherGiguere2006} and estimating the time needed to cross the forward shock. The injection spectrum is assumed to be a broken power law, with slope $\sim$1.5 up to an energy of $E_b \sim 500$~GeV and $\sim$2.4 at higher energies~\cite{Bykov2017}. The luminosity is determined by the initial rotation period $P_0$ for which we assume a Gaussian distribution centered at $\langle P_0 \rangle = 100$ msec with standard deviation $\sigma_{P0} = 50$~msec as in Ref.~\cite{Amato2018}.
By using Eq.~\ref{eq:burstsolution} for pulsars we are in fact assuming that we can approximate the injection from these sources as a burstlike event. That is, however, a good approximation as far as low energies are considered. 
In any case, the pulsar contribution is mostly to be used to determine the flux of positrons (and hence the few electrons that are contributed by pulsars). 
We found that in order to reproduce the date we need an efficiency of~$\epsilon_{\textrm{PWN}} = 15$\%.

Finally, we describe the injection and propagation of secondary leptons (produced by CR interactions in the ISM) by modeling the interaction with the ISM of a flux of protons and helium nuclei as measured by AMS-02~\cite{AMS02-H,AMS02-he}, assuming for the ISM the gas distribution as in Ref.~\cite{Strong1998}. 
We notice however that only secondary positrons at energies below $\sim 30$ GeV provide a sizable contribution to the observed fluxes, the secondary contribution to local electrons being negligible at all energies~\cite{Manconi2019,Fornieri2020}.

In Fig.~\ref{fig:spectra} we show the median of 1000 Monte Carlo realizations with the same source and propagation parameters. The uncertainty band shows the 2$\sigma$ fluctuation around the median due to the individual realization.
%
%
The spectra of electrons and positrons resulting from SNe, CR interactions in the Galaxy and pairs released by bow-shock PWNe are shown separately and their sum is compared with AMS-02 data~\cite{AMS02-electrons,AMS02-positrons}. 
We focused on $E \gtrsim 10$ GeV, so that the effects of solar modulation can be considered of little impact~\cite{AMS02-modulation}.

The fit to the observed electron spectrum requires that primary electrons (from SNRs) are injected with a slope $\gamma = 2.39$.
The propagated spectrum shown in the left-hand panel of Fig.~\ref{fig:spectra} clearly shows a prominent feature that starts around $\sim 40$ GeV and reproduces the data very nicely. This feature is solely due to the onset of the KN regime in ICS off the UV photons (see Fig.~\ref{fig:losses}). 

One might be tempted to attribute this feature to a combination of other effects, such as the contribution of pulsars and the change of slope in the diffusion coefficient (as in Refs.~\cite{Evoli2019,Evoli2020}), but we checked that this is not so. The fraction of electrons that is contributed by pulsars is severely constrained by the flux of positrons from PWNe as plotted in the right-hand panel of Fig.~\ref{fig:spectra}. If one subtracts the electrons of pulsar origin from the total electron flux obtains the dashed orange line in the right-hand panel. This curve shows the same feature very prominently, and only its normalization in the energy range $40\div1000$ GeV is reduced, by less than $\sim 20\%$. This fraction is exactly the contribution of pulsars to the electron flux at Earth, but it does not affect the presence of the feature. 

By the same token, the feature is unrelated to the change of slope in the diffusion coefficient, that is considered to be responsible for the hardening in the spectra of nuclei \cite{Genolini2017,Evoli2019}. In order to make this assessment we neglected the transition to KN in the ICS cross section and only included the change of slope of the diffusion coefficient (dotted red line in the right-hand panel of Fig.~\ref{fig:spectra}). No feature is visible in the electron spectrum in this case, thereby confirming once more that the feature in the electron spectrum is due to the fact that ICS off the UV photons has a transition from Thompson to KN regime in the energy region where $E\simeq \frac{m_{\rm e}^{2} c^4}{2k_{\rm B}T_{\rm UV}}\sim 50$ GeV. 
In fact the correction to the ICS cross section starts at somewhat lower energies and becomes evident already at $E\sim 40$ GeV, as illustrated in Fig.~\ref{fig:losses}. 

We conclude that the presence of this feature in the electron spectrum does not suggest the transition between two different types of sources of CR leptons, as advocated in Ref.~\cite{AMS02-electrons}, but rather a well-established phenomenon associated to electron energy losses. 
It is also worth stressing that the two-source approach discussed in Ref.~\cite{AMS02-electrons} would require that the lower energy contribution be dominated by a very steep spectrum with slope $\sim -4.31$, which is hard to reconcile with any kind of astrophysical accelerator, even after accounting for transport effects. 

{\it Conclusions.} We calculated the spectrum of electrons and positrons at Earth resulting from diffusive transport in the Galactic magnetic field under the action of radiative losses due to synchrotron emission and ICS on the CMB, IR, and UV photons in the ISM. Electrons are mainly primary particles resulting from acceleration in SNR shocks, and from pulsar winds in bow-shock nebulae. The latter also produce positrons, most likely responsible for the rising positron fraction. We describe both AMS-02 electron and positron spectra very nicely. In particular, we show that a feature appears in the electron spectrum as a result of the onset of KN effects in the cross section of ICS of electrons with UV photons in the ISM\footnote{Speculations about the transition to Klein-Nishina as a potential source of features in the electron spectrum, though at different energies, were already proposed in~\cite{Walt1991,Schlickeiser2010,Stawarz2010}.}. 
The feature starts at $\sim 40$ GeV and is most evident around $\sim 50$ GeV, corresponding to the energy where electrons scatter in the KN regime with the peak of the UV photon distribution. 

We exclude that the feature may be dominated by the electrons produced (together with an equal number of positrons) from pulsars. We also exclude the possibility that, at least partially, the feature may reflect the energy dependence of the diffusion coefficient, invoked to describe the spectral hardening in the spectra of nuclei. 

We also took into account the stochasticity in the spatial and temporal distribution of SN explosions and pulsars in the spiral arms of the Galaxy. The role of fluctuations on the spectrum of electrons and positrons is not significant at the energies where the feature is measured, while it becomes appreciable at energies larger than a few hundred GeV and eventually dominant at supra-TeV energies, although we do not discuss this regime here. 

In conclusion, the detection of the feature at $42.1^{+5.4}_{-5.2}$ GeV by AMS-02 shows in a rather clear way that the transport of electrons at such energies is loss dominated, thereby confirming independently that the size of the halo should be relatively large, as also found in analyses of the Be/B ratio \cite{Evoli2020}. These findings cast some doubts on the reliability of alternative models of CR transport developed in order to explain the positron spectrum as solely resulting from inelastic $pp$ collisions in the ISM~\cite{Cowsik2016,Lipari2017,Lipari2019}. Such models require that leptons' transport only becomes loss dominated at $E\gtrsim 300$ GeV.

{\it Acknowledgements.} The authors are very grateful to Paolo Lipari for numerous discussions on the topics of the present Letter. 
We acknowledge support by INAF and ASI through Grants No.~INAF-MAINSTREAM 2018 and ASI/INAF No.~2017-14- H.O and by the National Science Foundation under Grant No.~NSF PHY-1748958.

\bibliographystyle{myapsrev4-2}
\bibliography{mybibliography}

\end{document}